\documentclass[aps,prb,twocolumn,amsmath,amssymb,nofootinbib,superscriptaddress,floatfix]{revtex4-1}
\usepackage{amsmath}
\usepackage{amssymb}
\usepackage{graphicx,color}
\usepackage{color} 

\usepackage{bm} 
\usepackage{graphicx}
\usepackage{amssymb}
\usepackage{amsmath}
\usepackage{curves}
\usepackage{dcolumn} 

\usepackage{srcltx}
\usepackage[hypertex]{hyperref}

\begin{document}


\title{
Competing orders in the Dirac-like electronic structure
and 
the non-linear sigma model 
with the topological term
}

\author{Pouyan Ghaemi}
\affiliation{Department of Physics, University of California at
Berkeley, Berkeley, CA 94720, USA}
\affiliation{Materials Sciences Division, Lawrence Berkeley National Laboratory, Berkeley, CA 94720. USA}

\author{Shinsei Ryu}
\affiliation{Department of Physics, University of California at
Berkeley, Berkeley, CA 94720, USA}

\date{\today}

\begin{abstract}
The Dirac-like electronic structure can host a large number of
competing orders in the form of mass terms. In particular, two
different order parameters can be said to be dual to each other, 
when a static defect in one of them traps a quantum number (or ``charge'') of the other. 
We discuss that such complementary nature of the pair of the order parameters
shows up in their correlation functions and dynamical properties 
when a quantum phase transition is driven by fluctuations of the one of the order parameters.
Approaching the transition from the disordered (paramagnetic) side, the order parameter correlation
function at the critical point is reduced,
while such fluctuations enhance the correlation of the dual order parameter.
Such complementary behaviors in the correlation function can be used to
diagnose the nature of quantum fluctuations that is the driving force of
the quantum phase transition.
\end{abstract}

\maketitle

\newcommand{\Slash}[1]{\ooalign{\hfil/\hfil\crcr$#1$}}

\section{Introduction}
\label{sec: introduction}

Electrons in graphene
\cite{Novoselov05, Zhang05}, one atom-thick flake of carbon, 
effectively behave as relativistic particles, governed by the 
(2+1)-dimensional Dirac Hamiltonian:
the hopping of $\pi$ orbitals on the honeycomb network of carbon atoms
results in two Dirac points at the corners of the hexagonal Brillouin zone
($\boldsymbol{K}_{\pm}$). 
The Dirac electronic structure by now is not 
an exclusive feature of graphene, 
but also appears 
on the surface of the (3+1)-dimensional $\mathbb{Z}_2$ topological insulator
\cite{kanerev} 
such as 
Bi$_2$Se$_3$\cite{Hsieh08,Hsieh09} and 
Bi$_2$Te$_3$\cite{Xia09,Hsieh09b}, 
and in the quasi-two-dimensional organic conductor
$\alpha$-(BEDT-TTF)$_2$I$_{3}$.
\cite{Kobayashi2004, Katayama2006}

In the Dirac electronic structure, a mass gap can be induced in the fermionic energy spectrum 
by a non-vanishing and uniform order parameter.
In graphene, a variety of order parameters take on a form of such a mass term,
including, e.g.,  
the antiferromagnetism, valence bond solid (VBS), charge density wave (CDW), 
and superconductivity.
Some of these mass terms are of topological nature,
in the sense that when added
they realize a topological insulator, 
such as the quantum Hall effect (QHE) or
quantum spin Hall effect (QSHE),
or a topological superconductor. 
(See, for example, 
Refs.\ \onlinecite{Hou, Nomura, Semenof,Haldane88, Herbut07, ghaemi2009,ghaemi2007,santos,bergman},
and references in Ref.\ \onlinecite{Ryu09}).

Order parameters (mass terms) proximate to the Dirac electronic structure 
can be classified according to their algebraic properties. 
While they all anticommute with the Dirac kinetic term, 
as they induce a mass gap, 
they can either mutually commute or anticommute with each other 
(See below for more details).
A set of mass terms which anticommute with each other do not compete,
in the sense that one can adiabatically changes one order into another 
without closing a gap in the fermionic spectrum. 
Such mass terms can then be unified into a single multi-component 
order parameter,
and each mass can be viewed as representing a different direction 
in the order parameter space.
For example, the three antiferromagnetic orders (N\'eel orders) and 
the two different patterns of the VBS do not compete and,
at the level of the fermionic spectrum,
can be integrated into a five-component 
($\mathrm{O}(5)$) vector.

When a topological defect is introduced in such a multi-component 
order parameter,
midgap states appear in the massive Dirac energy spectrum.
For example, for the case of the N\'eel-VBS 5-tuplet, 
one can create a vortex, say, in the $x$ and $y$ components 
of the N\'eel vector,
or,
in the two components of the VBS order parameters.
A closer look at these topological defects and the midgap states
reveals an interesting ``duality'' relation: 
A topological defect in the VBS order parameter 
supports two midgap states which
carry a spin quantum number (i.e., form a doublet of $S=1/2$).  
On the other hand, a topological defect in the N\'eel vector,
again, supports two midgap states which
can be viewed as forming a doublet of the ``valley pseudospin'' 
degree of freedom.
Here, by ``valley pseudospin'', we mean 
the degree of freedom associated with the two Dirac cones (valleys) 
in the honeycomb lattice, 
and regard two valleys at $\boldsymbol{K}_{\pm}$ 
as ``pseudo spin up'' and ``pseudo spin down'', respectively.
Such duality relation among order parameters can be found
also in (3+1) dimensions. \cite{hosur}

So far, the order parameters have been treated as a static background;
they are either treated at a mean field level,
or induced externally.
Accordingly, 
topological defects are also introduced statically and by an external mean. 
In this paper, 
we turn 
our attention to 
the effects of electron correlations
and on the dynamics of such order parameters.

While the vanishing density of states in the Dirac spectrum
at the half-filling prevents order parameters from developing
long-range order, 
when electron-electron interactions, say, are sufficiently strong,
such orders can still be induced.

Here, an important implication of the duality relation between 
order parameters,
found solely at the level of the fermionic energy spectrum, 
is the possibility of ``deconfined'' quantum criticality:
I.e., 
a possible \textit{continuous} (second order) quantum phase transition
between symmetry unrelated quantum phases,
which is forbidden in the conventional Landau-Ginzburg framework.
\cite{SenthilScience}
The duality relation between different orders,
i.e., a topological defect in one of the two 
carry a quantum number (``charge'') of the other, is a prerequisite for 
deconfined quantum criticality;
with the duality, collapsing one phase by condensing defects
automatically leads to the emergence of the other phase.
I.e., they are connected through a continuous transition,
and do not coexist.

The deconfined quantum criticality scenario has been investigated so far 
in the square-lattice quantum antiferromagnet;
Indeed, a recent numerical study of the $J$-$Q$ model
\cite{sandvik,Kaul2008}
observed a vortex-like texture of the VBS order parameter
induced around a deficit of spin  
in the both columnar and plaquette VBS phases.

In this paper,  
including sufficiently strong short-range electron-electron interactions, 
we induce a mass gap in the Dirac spectrum dynamically (spontaneously), 
and study the structure of the resulting phase diagram and critical properties,
with eyes on to the scenario of deconfined quantum criticality.
In Sec.\ \ref{sec Dirac Hamiltonian and order parameters},
we introduce the Dirac Hamiltonian 
together with mass terms that represent order parameters of
various kinds. The four-fermion interactions which result
in a spontaneous generation of these order parameters are also
introduced. 
One of our main results is
the anomalous dimensions of the masses at the non-trivial
critical point, 
Eqs.\ (\ref{sum large N result}) and (\ref{sum large espilon result})
in Sec.\ \ref{Large-N and epsilon-expansions}.
The detailed derivation of these are presented 
in Subsecs.\ \ref{Large-Nf expansion}-\ref{Renormalization Conditions}. 
We conclude in Sec.\ \ref{Discussion} with our speculation
on the two-parameter renormalization group (RG) flow 
in the non-linear sigma model (NL$\sigma$M) 
with a topological term
($\theta$-term) 
in (2+1) dimensions.

\section{Dirac Hamiltonian and order parameters}
\label{sec Dirac Hamiltonian and order parameters}

We start by describing 
the Dirac kinetic term of graphene Hamiltonian,
which is represented by 
the single particle Hamiltonian $\mathcal{H}_0(\textbf{k})$
in momentum space as 
\begin{align}
\mathcal{H}_0(\textbf{k})
 = 
k_{x} \sigma_3 \otimes \tau_1
+
k_{y} \sigma_3 \otimes \tau_2, 
\end{align}
where $\sigma_{1,2,3}$ and $\tau_{1,2,3}$ are two independent 
sets of $2\times 2$ Pauli matrices,
and we have set the Fermi velocity to be unity, $v_F=1$. 
In graphene the Pauli matrices $\tau_{x,y,z}$ act on 
the valley index whereas the Pauli matrices $\sigma_{x,y,z}$ act on
the sublattice index. 

To describe several order parameters, 
including e.g., magnetic and superconducting order parameters, 
we introduce two more gradings, 
one for spin (represented by the Pauli matrices $s_{1,2,3}$),
and 
the other for particle-hole 
(represented by the Pauli matrices $\mu_{1,2,3}$). 
(For more detailed description of the order parameters
and the corresponding mass matrices entering in the Dirac Hamiltonian,
see Ref.\ \onlinecite{Ryu09}.)

\subsection{fermion mass terms in graphene}

We now introduce mass terms in the Dirac Hamiltonian
$\mathcal{H}_0({\bf k})$. 
To discuss all possible duality relations among order parameters
discussed in 
Ref.\ \onlinecite{Ryu09}
in a unified fashion,
let us start from a set of seven $2^{3}\times2^{3}$ anticommuting
hermitian matrices,
\begin{align}
&
\xi_{i}\xi_{j}+\xi_{j}\xi_{i}=2\delta_{ij},
\quad i,j=1,\ldots,7,
\nonumber \\
\mbox{with}
\quad
&
\xi_{7}= {i} \xi_{1}\xi_{2}\cdots\xi_{6}.
\end{align}
These matrices form a spinor representation of $\mathrm{SO}(7)$.
Two out of these seven matrices, $\xi_{1,2}$, can be used to form a
Dirac kinetic Hamiltonian, whereas the remaining five matrices can be used
as a mass matrix representing an order parameter: 
\begin{align}
\mathcal{H}(\textbf{k})
&=
\sum\nolimits _{i=1}^{2}k_{i}\xi_{i}
+\sum\nolimits _{i=3}^{7}m_{i}\xi_{i},
\nonumber \\
H
&=
\int d^2 k\, 
\Psi^{\dag}
\mathcal{H}(\textbf{k})
\Psi.
\end{align}
 Here $m_{i=3,\ldots,7}\in\mathbb{R}$ represents a five-component
order parameter;
$\Psi$ is a fermionic field operator
which includes sublattice, valley, spin, and particle-hole 
gradings. For the cases of our interest, 
it turns out we can always take 
$\Psi$ to be eight-component. 
\cite{Ryu09}
The imaginary-time Lagrangian is given by
\begin{align}
\mathcal{L}
 & =  
\Psi^{\dag}
\Big(\partial_{\tau}
+\sum\nolimits _{i=1}^{2}k_{i}\xi_{i}
+\sum\nolimits _{i=3}^{7}m_{i}\xi_{i}
\Big)
\Psi.
\end{align}

\subsection{four-fermion interactions}\label{ffermion}

So far, we have been discussing 
order parameters $m_i$ and their mass matrices $\xi_i$ ($i=3,\ldots,7$)
entering in the single-particle Dirac Hamiltonian, 
without asking their microscopic origin, i.e. 
microscopic interactions which can generate these orders 
dynamically. 
An interacting Hamiltonian which gives rise to these orders
is:
\begin{align}
H
&=
\int d^2 k\, 
\Psi^{\dag}
\left(
\xi_1 {k}_x
+
\xi_2 {k}_y
\right)
\Psi 
\nonumber \\
&\quad 
-
\int d^2 r\,  \sum^7_{a=3}
\frac{g_a}{2}
\left(
\Psi^{\dag}
\xi_a 
\Psi
\right)^2,
\label{eq: generic Hamiltonian}
\end{align}
where $g_a>0$. 
Such model can be designed on the lattice
with extended Hubbard type interactions. 
With 
the Hubbard-Stratonovich (HS) transformation, 
these quartic interactions can be decoupled into 
channels, such as
antiferromagnetic,
superconducting, 
and VBS orders that we have discussed. 
If the coupling constants $g_a>0$ are large enough, we find a saddle 
point where these order parameters are non-zero. 

While all coupling constants $g_a$ can take, in principle, different values,
below, let us first consider an interacting Hamiltonian
with $g_{a}=g$ ($a=3,\ldots,7$), 
\begin{align}
H
&=
\int d^2 k\, 
\Psi^{\dag}
\left(
\xi_1 {k}_x
+
\xi_2 {k}_y
\right)
\Psi
\nonumber \\
&\quad
-
\int d^2 r
\frac{g}{2}
\sum^7_{a=3}
\left(
\Psi^{\dag}
\xi_a 
\Psi
\right)^2.
\label{eq: O(5) Hamiltonian}
\end{align}
This is an $\mathrm{O}(5)$ symmetric analogue of
the (three-dimensional version of the) Gross-Neveu (GN) model.
\cite{Gross_Neveu1974}
The GN model with
$\mathrm{O}(1)\simeq \mathbb{Z}_2$ internal symmetry
and
the chiral GN model with
$\mathrm{O}(2)\simeq \mathrm{U}(1)$ internal symmetry
have been studied extensively,
as a prototype model to discuss spontaneous chiral symmetry breaking. 
The corresponding Euclidean Lagrangian is
\begin{align}
\mathcal{L}
=
\Psi^{\dag}
\left(
\partial_{\tau}
+
\xi_1 {k}_x
+
\xi_2 {k}_y
\right)
\Psi
-
\frac{g}{2}
\sum^7_{a=3}
\left(
\Psi^{\dag}
\xi_a 
\Psi
\right)^2. 
\end{align}

We can successively break this 
$\mathrm{O}(5)$ symmetry
down to $\mathrm{O}(4)$, 
to $\mathrm{O}(3)$, 
to $\mathrm{O}(2)\times \mathrm{O}(2)$, 
etc. 
In the following,
in addition to the $\mathrm{O}(5)$ symmetric model, 
we consider
$\mathrm{O}(4)$ 
and
$\mathrm{O}(3)$ 
symmetric models as well. 
These models can be obtained from 
Eq.\ (\ref{eq: generic Hamiltonian}) by setting
$(g_3, g_{4,5,6,7})=(0,g)$ (for the $\mathrm{O}(4)$ symmetric model)
or 
$(g_{3,4}, g_{5,6,7})=(0,g)$ (for the $\mathrm{O}(3)$ symmetric model).
In these cases, 
we can treat the remaining $\xi$ matrices (i.e. $\xi_3$ for the $\mathrm{O}(4)$ symmetric model or $\xi_3$ and $\xi_4$ for the $\mathrm{O}(3)$ symmetric model) 
as mass terms. 
Schematically, 
\begin{align}
H
&=
\int d^2 k\, 
\Psi^{\dag}
\Big(
\xi_1 {k}_x
+
\xi_2 {k}_y
+
\sum_{b \in M} m_b \xi_b
\Big)
\Psi
\nonumber \\
&\quad 
\qquad 
-
\frac{g}{2}
\int d^2 r\,
\sum_{a \in \bar{M}}
\left(
\Psi^{\dag}
\xi_a 
\Psi
\right)^2,
\end{align}
where 
$M=\emptyset, \{3\},\{3,4\}$,
for the O(5), O(4), and O(3) symmetric models,
respectively,
and 
$\bar{M}$ is the complement of $M$: 
We have added mass terms
$\sum_{b \in M} m_b \xi_b$
which are not generated spontaneously by the interactions. 
Rather, they are entering here as a parameter that changes 
the band structure of the single particle Hamiltonian. 
We will explore the phase diagram in terms of the coupling
constant $g$ as well as the masses $m_b$ below.

When the coupling constant(s) is(are) large enough, 
the system is in an ordered phase. 
The phase diagram can be explored by 
saddle-point or mean field approximation.
With the HS transformation,
\begin{align}
\mathcal{L}
&=
\Psi^{\dag}
\left(
\partial_{\tau}
+
\xi_1 {k}_x
+
\xi_2 {k}_y
+
\sum\nolimits_b m_b \xi_b
+
\sum\nolimits_a v_a \xi_a 
\right)
\Psi
\nonumber \\
&\quad 
+
\frac{1}{2g}
\sum\nolimits_a v^2_a.
\end{align} 
If we freeze the dynamics of the HS fields ($v_a$), we get the meanfield Hamiltonian 
discussed before. In the following, however, we are interested in 
the case where the HS fields are dynamical. 

The 
$\mathrm{O}(N_{\Sigma})$ ($N_{\Sigma}=3,4,5$ is the number of $\Sigma_a$ matrices introduced below)
symmetry of the problem can be seen 
most clearly if we make the following change of basis
in the fermionic path integral variables:
\begin{eqnarray}
&&
\gamma_0 := -
\xi_{3}
\xi_{4}
\xi_{5}
\xi_{6}
\xi_{7},
\quad
\gamma_1 :=  -{i}  \gamma_0 \xi_1,
\quad
\gamma_2 :=  -{i}  \gamma_0 \xi_2,
\nonumber \\
&&
\Sigma_{a}
:=
\gamma_0
 \xi_{a+2},
\quad (a=1,\ldots,N_{\Sigma}),
\nonumber \\
&&
\bar{\psi} :=\Psi^{\dag} \gamma_0, 
\quad
\psi:=\Psi,
\label{new basis}
\end{eqnarray}
wherein the Lagrangian in terms of the new variables is given by:
\begin{align}
\mathcal{L}
 & = 
\bar{\psi}
\Big(
\partial_{\mu}\gamma_{\mu}
+\sum\nolimits_{a=1}^{N_{\Sigma}} v_{a} \Sigma_{a}
\Big)\psi+
\frac{1}{2g}
\sum\nolimits_{a=1}^{N_{\Sigma}} v^2_a. 
\label{eq: canonical lagrangian}
\end{align}
Here, the summation over space-time index $\mu=0,1,2$
is implicit, and we have switched off the masses $m_b$
($b\in M$)
momentarily. 
The merit of this change of variables is
that it untangles rotations in the order parameter space
and in the real space:
the mass matrices ($\Sigma_{a=1,\ldots,N_{\Sigma}}$)
and the matrices entering in the Dirac kinetic term 
($\gamma_{\mu=0,\ldots,2}$)
are made mutually commuting,
\begin{align}
\left[\gamma_{\mu},\Sigma_{a}\right]
=0,\quad\forall\mu,a,
\end{align}
where $\gamma$s and $\Sigma$s form 
$\mathrm{SO}(3)$ and $\mathrm{SO}(N_{\Sigma})$,
respectively, 
\begin{eqnarray}
 &  & 
\gamma_{\mu}\gamma_{\nu}
+\gamma_{\nu}\gamma_{\mu}=
2\delta_{\mu\nu},\quad\mu,\nu=0,1,2,\nonumber \\
 &  & 
\Sigma_{a}\Sigma_{b}+\Sigma_{b}\Sigma_{a}=2\delta_{ab},
\quad a,b=1,\ldots,N_{\Sigma}.
\end{eqnarray}
The $\mathrm{O}(N_{\Sigma})$ rotation acting on 
the fermion fields $\psi$ can be generated by
unitary matrices
$
\exp\left[
{i} X^{ab}\right]
$
where
$X^{ab}=
\left[
\Sigma_a, \Sigma_b
\right]
$.
With simultaneously rotating 
the $N_{\Sigma}$-component vector $v_a$ 
appropriately, 
the Lagrangian is left invariant. 
Below, 
we will denote 
the dimension of 
$\Sigma$s and $\gamma$s 
by $D_\Sigma$ and $D_\gamma$, respectively.
In the above construction,
$D_\Sigma=D_\gamma=8$,
while our calculations presented below are valid for any
$D_\Sigma$ and $D_\gamma$. 

\subsection{meanfield phase diagram}

To control
the saddle point approximation
(the mean field theory), 
and to prepare for the subsequent large-$N_f$ expansion, 
we generalize our Lagrangian to include $N_f$ flavors of fermions.
In the imaginary time formalism (the Euclidean signature),
the large-$N_f$ generalized Lagrangian is then given by 
\begin{align}
\mathcal{L}
&=
\sum^{N_f}_{\iota=1}
\bar{\psi}_{\iota}
\left(
\gamma_{\mu} \partial_{\mu}
+
\sum\nolimits_a v_a \Sigma_a 
+
\sum\nolimits_b m_b \Sigma_b 
\right)
\psi_{\iota}
\nonumber \\
&\quad 
+
\frac{1}{2g}
\sum\nolimits_a v^2_a, 
\label{lnf}
\end{align}
where $\gamma_{\mu}$s are Euclidean gamma matrices. 

We now look for a spatially homogeneous
and 
$\mathrm{O}(N_{\Sigma})$ symmetric saddle point solution
by setting 
\begin{eqnarray}
\boldsymbol{v}(x) = |v| \hat{\boldsymbol{n}},
\quad
\hat{\boldsymbol{n}}
\cdot
\hat{\boldsymbol{n}}=1. 
\label{nvector}
\end{eqnarray}
When $m_b=0$ the self-consistency condition for $|v|$ is given by
\begin{align}
&
D_{\gamma}
 |v|
\int^{\Lambda} \frac{d^3 k}{(2\pi)^3}
\frac{1}{k^2+|v|^2}
-
\frac{|v|}{2N_f g}
=
0.
\end{align}
Here we have introduced
the ultra-violet (UV) cutoff $\Lambda$ by hand.

\begin{figure}[t] 
\begin{center}
\unitlength=10mm
\begin{picture}(6,4)(-3,-0.5)

\thinlines



\put(-3,0){\vector(1,0){6.5}}
\put(0,-0.5){\vector(0,1){4}}

\put(0,0){\line(0,1){1}}

\put(0,1.1){\vector(0,-1){0.5}}
\put(0,0.6){\vector(0,-1){0.5}}

\put(0,0)  {\vector(1,0){0.7}}
\put(0.7,0){\vector(1,0){0.7}}

\put(0,0)    {\vector(-1,0){0.7}}
\put(-0.7,0){\vector(-1,0){0.7}}

\thicklines

\put(-1,3.2){$g N_f \Lambda$}
\put(2.7,0.2){$m_b/\Lambda$}
\put(-0.25,-0.25){$0$}

\put(0.7,0.7){disordered}

\put(-2.3,0.7){disordered}

\put(-1,2){$\mathrm{O}(N_{\Sigma})$ ordered}

\curve(
	 0.000000, 1.000000,
	 0.500000, 1.090909,
	 1.000000, 1.200000,
	 1.500000, 1.333333,
	 2.000000, 1.500000,
	 2.500000, 1.714286,
	 3.000000, 2.000000)

\curve(
	 -3.000000, 2.000000,
	 -2.500000, 1.714286,
	 -2.000000, 1.500000,
	 -1.500000, 1.333333,
	 -1.000000, 1.200000,
	 -0.500000, 1.090909,
	 0.000000, 1.000000)

\end{picture}
\end{center}
\caption{
\label{fig: phase dirgram}
The mean field phase diagram near the Dirac point.
        }
\end{figure}
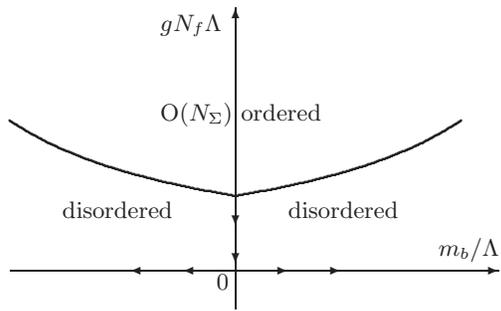

The resulting mean-field 
phase diagram is depicted in Fig.\ \ref{fig: phase dirgram},
wherein we also include 
$m_{b}$ $(b\in M)$: 
For sufficiently small $g \ll 1/(N_f \Lambda)$, 
there are two disordered phases where fermion fields
are completely gapped by the mass $m_{b\in M}\neq 0$.
They are separated by a phase boundary $m_b=0$.
In particular, when $m_{b}$ represents 
the Kane-Mele mass term (QSHE mass term)
it is a quantum phase transition separating
the trivial band insulator
and the topological (QSH) insulator. 
For sufficiently large $g \gg 1/(N_f \Lambda)$, 
$\mathrm{O}(N_{\Sigma})$ symmetry is spontaneously broken. 
The arrows in Fig.\ \ref{fig: phase dirgram}
indicates the infra-red (IR) renormalization group (RG) flow.
Just at the non-interacting Dirac point, 
four fermion interactions are irrelevant,
whereas the mass terms are relevant,
from the power-counting.
The transition between two
disordered phases is then described 
by the non-interacting Dirac point. 
The nature of the phase boundary between the ordered phase
and disordered phases is more difficult to discuss. 
The critical point at $m_b=0$ with $gN_f \Lambda\neq 0$
can be nevertheless accessible with the large-$N_f$ 
or $\epsilon$ expansion ($4-\epsilon$ expansion),
as we explore in the next section.

\section{Large-$N_f$ and $\epsilon$ expansions}
\label{Large-N and epsilon-expansions}

\subsection{summary of results} 

In this section, we focus on the nature of the critical point
separating the ordered phase and the semi-metallic Dirac phase
located at $m_b=0$ with $gN_f \Lambda\neq 0$.
This IR unstable fixed point will be called the GN fixed point.
While it is not perturbatively accessible in the coupling constant $g$,
it can be studied either by the large-$N_f$ expansion
or by the $\epsilon$ expansion. 
\cite{kocic,moshe,justinbook}
The problem of the (2+1)-dimensional Dirac fermions 
interacting via short-range as well as long-range 
(Coulombic as well as gauge) interactions has a long history; 
in particular, the recent fabrication of  graphene, 
and in particular
suspended graphene \cite{andrei},
motivated many papers to revisit this problem
(see for example, Ref.\ \onlinecite{review}).

One of our main focuses in this paper is 
the effects of four-fermion interactions
on the duality relation among order parameters. 
Of particular interest are the scaling dimensions 
of the mass bilinears at the GN fixed point
associated to the order parameters;
we can compare the scaling dimensions of 
the mass $\bar{\psi}\psi$
and 
the mass $\bar{\psi}\Sigma \psi$, 
where the mass matrix $\Sigma$ 
corresponds 
to one of the masses described by $m_b$ ($\Sigma \in \gamma^0 \xi_b$ with $b\in {M}$),
and $\Sigma \notin \{\Sigma_{a}\}_{a=1,\ldots, N_{\Sigma}}$. 
I.e., $\Sigma$ is an identity 
matrix in the Dirac indices, and anticommutes with
$\Sigma_{a}$ for all $a=1,\ldots, N_{\Sigma}$
(i.e., the mass $\bar{\psi}\Sigma \psi$ is dual to 
$\bar{\psi}\Sigma_{a}\psi$),
whereas 
the mass $\bar{\psi}\psi$ commutes with $\bar{\psi}\Sigma_a\psi$.

To leading order in the large-$N_f$ expansion, 
the anomalous dimension of 
the fermion field $\psi$ ($\eta_{\psi}$),
the mass bilinear $\bar{\psi}\psi$ ($\eta_{\bar{\psi}\psi}$),
and the mass bilinear $\bar{\psi}\Sigma \psi$ ($\eta_{\bar{\psi}\Sigma \psi}$),
are given by
\begin{align}
\eta_{\psi}
&=
\frac{8\, N_{\Sigma}}
{ 3 \pi^2 \tilde{N} }, 
\nonumber \\
\eta_{\bar{\psi}\psi}
&=
\frac{16\,  N_{\Sigma} }{3 \pi^2\tilde{N}}
(1+3)=
\frac{64\,  N_{\Sigma} }{3\pi^2 \tilde{N}}, 
\nonumber \\
\eta_{\bar{\psi}\Sigma \psi}
&=
\frac{16\,  N_{\Sigma} }{3\pi^2 \tilde{N}}
(1-3)=
-\frac{32\,  N_{\Sigma} }{3\pi^2 \tilde{N}},
\label{sum large N result}
\end{align}
where we have introduced 
\begin{align}
\tilde{N}
:=
D_\gamma D_\Sigma N_f.
\end{align}
Alternatively, we can also look at 
the anomalous dimensions
using the $\epsilon$ expansion
\cite{balents,ghaemi2005} 
(i.e., 
in $d=4-\epsilon$ dimensions where physical dimension 
corresponds to $\epsilon=1$). 
To leading order in $\epsilon$,
\begin{align}
\eta_\psi &= 
\frac{1}{2}\frac{\epsilon\,  N_\Sigma}{\tilde{N}+4-N_{\Sigma} },
\nonumber \\
\eta_{\bar{\psi}\psi} &= \frac{ \epsilon\, N_\Sigma}
{\tilde{N}+4-N_\Sigma}(1+2)=
\frac{3 \epsilon\, N_\Sigma}
{\tilde{N}+4-N_\Sigma}, 
\nonumber \\
\eta_{\bar{\psi}\Sigma \psi} &= \frac{\epsilon\, N_\Sigma}
{\tilde{N}+4-N_\Sigma}(1-2)=
\frac{-\epsilon\, N_\Sigma}
{\tilde{N}+4-N_\Sigma}.
\label{sum large espilon result}
\end{align}
The calculational details 
of
Eqs.\ (\ref{sum large espilon result})
and
(\ref{sum large N result})
can be found in 
the next subsection.

While the results in the large-$N_f$ and the $\epsilon$ expansions
do not match numerically,
the signs of the anomalous dimensions agree in both calculations.
In particular observe that $\eta_{\bar{\psi}\psi}$
and $\eta_{\bar{\psi}\Sigma \psi}$
have opposite sign; 
the positive/negative sign in these anomalous dimensions
directly comes from 
anticommutation/commutation relations 
of the masses and order parameters. 
I.e., the positive/negative sign is a direct manifestation of the duality. 
To be more precise, observed that in 
$\eta_{\bar{\psi}\psi}$ and $\eta_{\bar{\psi}\Sigma \psi}$ above, 
two contributions are separately displayed
(as seen in ``$(1\pm 3)$'' and ``$(1\pm 2)$''
in Eqs.\ (\ref{sum large espilon result}) and
(\ref{sum large N result}), respectively). 
The first terms in $\eta_{\bar{\psi}\psi}$ and $\eta_{\bar{\psi}\Sigma \psi}$ 
come from the renormalization of the single fermion field
($=2\eta_{\psi}$)
and do not know such duality relation,
while the second terms have opposite sign  (``$\pm 3$'' or ``$\pm 2$'')
for $\eta_{\bar{\psi}\psi}$ and $\eta_{\bar{\psi}\Sigma \psi}$
which originates from the  
anticommutation/commutation relations 
of the masses and order parameters. 
Such anticommutation/commutation relations also control 
presence/absence of topological terms 
when one integrates the fermionic degrees of freedom to derive the 
non-linear sigma model (NL$\sigma$M) (see below).

\subsection{Large-$N_f$ expansion}
\label{Large-Nf expansion}

We now present some calculational details of 
the large-$N_f$ and 
$\epsilon$-expansions.
In 
Secs.\ \ref{Large-Nf expansion}
and \ref{epsilon-expansion}, 
we first collect diagrammatic calculations for renormalization
constants,
in the large-$N_f$
and 
$\epsilon$-expansions, respectively. 
These data will be used in 
Subsec.\ \ref{Renormalization Conditions}
to compute anomalous dimensions of 
the fermion field and 
mass bilinears.

The bare Euclidean action 
of the $\mathrm{O}(N_{\Sigma})$ GN model
in terms of 
the fermionic ($\psi$) and bosonic fields ($v_a$) is
Eq.\ (\ref{lnf}): 
\begin{equation}
\mathcal{L}
=
\sum^{N_f}_{\iota=1}
\bar{\psi}_{\iota}
\Big(
\gamma_{\mu} \partial_{\mu}
+ 
\sum\nolimits^{N_\Sigma}_{a=1} v_a \Sigma_a 
\Big)
\psi_{\iota}
+
\frac{1}{2g}
\sum^{N_\Sigma}_{a=1} v^2_a, 
\label{lnf2}
\end{equation}
where we work in $d=3$ space-time dimensions,
$\mu=1,\ldots, d$.

\paragraph{Boson self-energy}

\begin{figure}[t]
\centering
\includegraphics[width=0.45\textwidth]{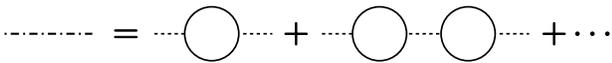}
\caption{
The boson self-energy 
in the large-$N_f$ expansion obtained
by summing fermion bubbles,
where 
full lines represent the fermion propagator,
and 
broken lines represent the boson propagator.
        } 
\label{fig:idia1}
\end{figure}

To leading order in the large-$N_f$ expansion,
the boson ($v_a$ field) self-energy
is obtained by summing 
over fermionic bubble diagrams (Fig.\ \ref{fig:idia1}),
leading to the boson propagator
\begin{align}
\langle v^*_a(\textbf{k}) v_b(\textbf{k}') \rangle 
=
\frac{16}{\tilde{N}}
\frac{\delta_{ab} \  \delta_{\textbf{k},\textbf{k}'}}{ |k|},
\label{boson propagator large N}
\end{align}
at the non-trivial critical point $g=g_*$.

\begin{figure}[htp]
\includegraphics[height=1.5cm]{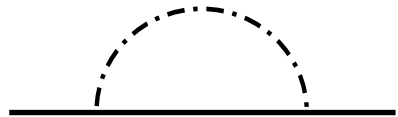}
\caption{The fermion self-energy diagram
where 
the broken line represents the boson propagator,
Eq.\ (\ref{boson propagator large N}).
\label{fig:idia2}
} 
\end{figure}

\begin{figure}[t]
\includegraphics[height=2.2cm]{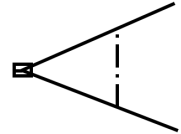}
\caption{
\label{fig vtx correction}
The vertex correction diagram relevant 
for the renormalization of
the mass bilinear 
$\bar{\psi} \psi$
or
$\bar{\psi} \Sigma \psi$,
where the square represents 
either
the insertion of the operator $\nu \bar{\psi} \psi$
or
$\rho \bar{\psi} \Sigma \psi$.
}
\end{figure}

To compute 
the anomalous dimensions of the fermionic field,
as well as 
the anomalous dimensions of mass terms of different kinds, 
at
the non-trivial critical point,
there are three diagrams to be considered:
Figs.\ \ref{fig:idia2}-\ref{fig vtx correction}.

\paragraph{Fermion self-energy}

The diagram corresponding to the Fermion self-energy 
is shown in Fig.\ \ref{fig:idia2}.
The corresponding integral is:
\begin{align}
&\quad 
\langle \bar{\psi}(\textbf{q})\psi(\textbf{q}) \rangle^{-1} 
\nonumber \\
&=
i \Slash{q}
+\int \frac{d^3 k}{(2\pi)^3} 
\frac{
i \Slash{k}
}{k^2}
\frac{16 \Sigma_a\delta_{ab}\Sigma_b}{\tilde{N} |\textbf{q}-\textbf{k}|} 
\nonumber \\
&=
i \Slash{q}\left[1+\Sigma^{N_f}_\psi 
\log\left(\frac{\Lambda}{|q|}\right)\right], 
\label{eq: fermion self large N}
\end{align}
where $\Slash{k}= k_{\mu}\gamma_{\mu}$, 
and
$\Sigma^{N_f}_\psi=
16N_\Sigma/(6\pi^2\tilde{N})$.
In Eq.\ (\ref{eq: fermion self large N}),
and from now on,
we suppress the flavor index, 
since correlation functions 
that consists of several fermionic operators 
are non-zero only when
all fermion fields carry the same flavor index:
E.g., 
$
\langle \bar{\psi}(\textbf{q})\psi(\textbf{q}) \rangle^{-1} 
=
\langle \bar{\psi}_{\iota}(\textbf{q})\psi_{\iota}(\textbf{q}) \rangle^{-1}
$
in Eq.\ (\ref{eq: fermion self large N}).

\paragraph{$\bar{\psi} \psi$ bilinear vertex correction}

To compute the anomalous dimension of the fermion bilinear
$\bar{\psi} \psi$,
we add a mass bilinear term $\nu \bar{\psi} \psi$ to the action,
where $\nu$ is the corresponding ``mass''. 
We also introduce a multiplicative renormalization constant for $\nu$,
$\Delta^{N_f}_\nu$, in the large-$N_f$ expansion. 
The vertex diagram corresponding to the renormalization of 
the mass $\nu$ is
Fig.\ \ref{fig vtx correction}: 
\begin{align}
&\quad 
\Delta^{N_f}_\nu \nu 
\log\left(\frac{\Lambda}{k}\right) 
\nonumber \\
&=
\nu \int \frac{d^3 q}{(2\pi)^3} 
\frac{i(\Slash{q}+\Slash{k})}{(q+k)^2}
\frac{16 \Sigma_a \delta_{ab} \Sigma_b}{\tilde{N}q}
\frac{i(\Slash{q}-\Slash{k})}{(q-k)^2} 
\nonumber \\
&=
-\nu
\frac{8 N_\Sigma}{\pi^2 \tilde{N} }
\log\left(\frac{\Lambda}{k}\right).
\end{align}


\paragraph{$\bar{\psi}\Sigma \psi$ bilinear vertex correction}
 
We now add a term of the form $\rho\bar{\psi}\Sigma\psi$
to the action.
Here, the mass matrix $\Sigma$ 
corresponds to one of the masses described by $m_b$ (i.e. $\Sigma \in \gamma^0 \xi_b$ with $b\in {M}$),
and $\Sigma \notin \{\Sigma_{a}\}_{a=1,\ldots, N_{\Sigma}}$:
$\Sigma$ is an identity 
matrix in the Dirac indices, and anticommutes with
$\Sigma_{a}$ for all $a=1,\ldots, N_{\Sigma}$.
The diagram that determines 
the renormalization of $\rho\bar{\psi}\Sigma\psi$
is similar to the one given above 
(Fig.\ \ref{fig vtx correction})
except for the presence of $\Sigma$ matrix:
\begin{align}
&\quad 
\Delta^{N_f}_\rho \rho \, 
\log\left(\frac{\Lambda}{k}\right) \Sigma  
\nonumber \\
&=
\rho  \int \frac{d^3q}{(2\pi)^3} 
\frac{i(\Slash{q}+\Slash{k})}{(q+k)^2} 
\frac{16 \delta_{ab}\Sigma_a \Sigma \Sigma_b}
{\tilde{N} q}\frac{i(\Slash{q}-\Slash{k})}{(q-k)^2} 
\nonumber \\
&=
\rho  
\frac{8 N_\Sigma}{\pi^2 \tilde{N}}
\log\left(\frac{\Lambda}{k}\right)  
\Sigma.
\end{align}

\subsection{$\epsilon$ expansion}
\label{epsilon-expansion}

The bare Euclidean action 
$S = \int d^d r\,\mathcal{L}$
of the 
Gross-Neveu-Yukawa model,
in terms of 
the fermionic ($\psi$) and bosonic fields ($v_a$),
is given by 
\begin{align}
\mathcal{L}
&=
\sum^{N_f}_{\iota=1}
\bar{\psi}_{\iota}
\big(
 \gamma_\mu\partial_\mu 
 + \lambda  \textit{v}_a \Sigma_a 
\big)
\psi_{\iota}
\nonumber \\
&
\quad 
+\frac{1}{2}  
\sum^{N_{\Sigma}}_{a=1}
\left[(\partial_\mu \textit{v}_a)^2+r(\textit{v}_a)^2\right] 
+u \left[\sum^{N_{\Sigma}}_{a=1} (\textit{v}_a)^2\right]^2.
\end{align}
We now work in $d=4-\epsilon$ space-time dimensions. 
The bare propagator for the bosonic field $v_a$ is 
\begin{equation}
\langle v^*_a(\textbf{k}) v_b(\textbf{k}') \rangle
=\frac{\delta_{ab}\delta_{\textbf{k},\textbf{k}'}}
{|\textbf{k}|^2+r}. 
\end{equation}

To one-loop, we need to consider,
for the part of the theory that involves only the bosonic field ($v_a$), 
the diagrams that appear in renormalizing the $\phi^4$ theory.
In addition, 
the vertex correction for coupling $\lambda$,
and the boson self energy correction by a fermion loop
need to be calculated.

\paragraph{Vertex correction for coupling $\lambda$}
\begin{figure}[tp]
\includegraphics[height=2.2cm]{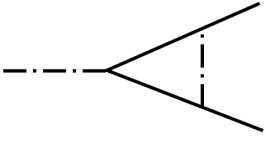}
\caption{The vertex diagram for $\lambda$ coupling.}
\end{figure}

The vertex correction of the coupling of 
the fermionic and bosonic fields $\lambda$ is given by the following diagram:
\begin{align}
&\quad 
\Delta^{\epsilon}_\lambda \lambda \Sigma_a  \frac{1}{\epsilon} 
\nonumber \\
& =
\lambda^3\int 
\frac{d^dq}{(2\pi)^d} 
\frac{
\Sigma_b 
{i} \Slash{q}
\Sigma_a 
i(\Slash{q}-\Slash{k})
\Sigma_b}  {\left[(q+p_2)^2+m^2\right]q^2(q-k)^2} 
\nonumber \\
& =
-\frac{2\lambda^3}{(4\pi)^2} 
\Sigma_a (-N_\Sigma+2) \frac{1}{\epsilon}. 
\end{align}

\paragraph{Boson self energy}

The one-loop boson self energy including a fermion loop
is given by 
\begin{align}
&\quad 
\langle 
v^*_a(\textbf{p}) 
v_b(\textbf{p}) \rangle^{-1} 
\nonumber \\
& = 
\delta_{ab} (\textbf{p}^2+r)
\nonumber \\
&
\qquad 
+\lambda^2 N_f \int \frac{d^dq}{(2\pi)^d} 
\mathrm{tr}\, \left[\Sigma_a\frac{
i \Slash{q}
}{\textbf{q}^2}\Sigma_b\frac{
i (\Slash{q}+\Slash{p})
}{(\textbf{q}+\textbf{p})^2}\right] 
\nonumber \\
&=
\delta_{ab} (\textbf{p}^2+r)+ 
\tilde{N}
\frac{p^2}{2}\int \frac{d^d q}{(2\pi)^d}\frac{\delta_{ab}}{\textbf{q}^2(q+p)^2}
\nonumber \\
&=
\delta_{ab} (\textbf{p}^2+r)+
\frac{\tilde{N}}
{(4\pi)^2} \frac{\textbf{p}^2}{\epsilon}\delta_{ab}.
\end{align}

\paragraph{Fermion self-energy}

The diagram corresponding to the Fermion self-energy 
has the same structure as in the large $N_f$
expansion (Fig.\ \ref{fig:idia1}) and 
the corresponding integral is given by 
\begin{align}
&\quad
\langle \bar{\psi}(\textbf{k}) \psi(\textbf{k}) \rangle^{-1} 
\nonumber \\
&=
i \Slash{k}
+ \lambda^2 \int \frac{d^d q}{(2\pi)^d}   
\frac{
i \Slash{q}
}{q^2}\frac{\Sigma_a\delta_{ab}\Sigma^b }{(\textbf{k}-\textbf{q})^2+r} 
\nonumber \\
&=
i\Slash{k}
\left(1+\Sigma_\psi^\epsilon\frac{1}{\epsilon}\right), 
\end{align}
where 
$\Sigma_\psi^\epsilon=\lambda^2 
N_\Sigma/(4\pi)^2$.

From these three diagrams, 
we read off the renormalization conditions 
for
the wavefunction renormalizations 
$Z^{\epsilon}_{\psi}$ 
and
$Z^{\epsilon}_{v}$
for the fermionic ($\psi$) and bosonic ($v_a$) fields, 
respectively, 
and for the coupling $\lambda$ as
\begin{align}
& 
Z^{\epsilon}_{\psi} i\Slash{p} + \lambda^2 \frac{N_\Sigma}{(4\pi)^2} i\Slash{p} 
= finite, 
\\
&  
Z^{\epsilon}_v
\textbf{p}^2+\textbf{p}^2\frac{1}{\epsilon}
\frac{\tilde{N}}{(4\pi)^2} 
= finite,  
\label{lv}
\\
&  
m^{\frac{-\epsilon}{2}}\lambda_0 
(Z^{\epsilon}_{v})^{\frac{1}{2}} 
Z^{\epsilon}_\psi-\frac{2\lambda^3}{(4\pi)^2}\frac{1}{\epsilon}  (-N_\Sigma+2) = \lambda, 
\label{lc}
\end{align}
where we have introduced 
an arbitrary mass scale $m\propto \sqrt{r}$,
and
$\lambda_0$ is the bare coupling which does not flow with
the mass scale. 
With the minimal subtraction scheme \cite{justinbook},
these conditions
can be used to derive, to one-loop,  
$Z^{\epsilon}_v$ and the beta function for $\lambda$, 
\begin{align}
Z^{\epsilon}_v
&=
1-
\frac{1}{\epsilon}\frac{\tilde{N}}{(4\pi)^2},
\nonumber \\
\beta_{\lambda^2}
&:=
-m\frac{\partial}{\partial m} \lambda^2
\\
&=
-\epsilon \lambda^2 
+ \frac{2\lambda^4}
{(4\pi)^2\epsilon}\left(\tilde{N}/2+4-N_\Sigma\right),
\nonumber
\end{align}
and also the critical coupling
\begin{eqnarray}
\lambda^{*2}= 
\frac{8\pi^2\epsilon}{\tilde{N}+4-N_\Sigma}. 
\end{eqnarray}

Below, we will compute the anomalous dimensions 
of the fermion field and the mass bilinears
at the non-trivial critical coupling 
$(\lambda, u)
=
(\lambda^*, u^*)
\neq (0,0)$. 
This fixed point in
the Gross-Neveu-Yukawa model is smoothly connected 
to the GN fixed point.
\cite{moshe,justinbook}
As in the large-$N_f$ expansion, 
we will consider the following three diagrams.

\paragraph{$\bar{\psi} \psi$ bilinear vertex correction}

To compute the anomalous dimensions,
we add the mass term $\nu \bar{\psi}\psi$ to the action,
which leads to 
the vertex diagram of type 
shown in Fig.\ \ref{fig vtx correction}. 
It determines, 
to leading order in $\epsilon$,
the 
renormalization constant 
$\Delta^\epsilon_\nu$
for 
the mass term $\nu \bar{\psi}\psi$
as
\begin{align}
\Delta^\epsilon_\nu \nu \frac{1}{\epsilon} 
&= 
\nu\lambda^2
\int \frac{d^dq}{(2\pi)^d} 
\frac{\Sigma_a \delta_{ab} i\Slash{q}  i\Slash{q} \Sigma_b}  
{\left[(\textbf{q}+\textbf{k})^2+m^2\right]q^2(q)^2} 
\nonumber \\
&=-\nu\frac{2\lambda^2 N_\Sigma}{(4\pi)^2}\frac{1}{\epsilon}. 
\end{align}

\paragraph{$\bar{\psi} \Sigma \psi$ bilinear vertex correction}
Similarly, 
by adding the mass term $\rho\bar{\psi} \Sigma \psi$ to the action,
the 
renormalization constant 
$\Delta^\epsilon_{\rho}$
for 
the mass term $\bar{\psi}\Sigma \psi$
is determined from 
the vertex diagram of type Fig.\ \ref{fig vtx correction}:
\begin{equation}
\begin{split}
\Delta^{\epsilon}_\rho \rho \Sigma \frac{1}{\epsilon}
&=\nu \lambda^2\int \frac{d^dq}{(2\pi)^d} 
\frac{\Sigma_a \delta_{ab} 
i \Slash{q}
\Sigma 
i \Slash{q}
\Sigma_b}  {\left[(q+p)^2+m^2\right]q^2(q)^2} 
\\ &=-\rho \frac{2\lambda^2}{(4\pi)^2}\frac{1}{\epsilon}
\Sigma_b \Sigma \Sigma_b=\frac{2\lambda^2}{(4\pi)^2}N_\Sigma \rho \Sigma \frac{1}{\epsilon}.
\end{split}
\end{equation}

\subsection{Renormalization conditions
and anomalous dimensions}
\label{Renormalization Conditions}

\begin{table}[t]
\centering
\begin{ruledtabular}
\begin{tabular}{ c  c  c  }
\ & large-$N_f$  expansion  & $\epsilon$ expansion   \\
\hline
\\
$
\displaystyle 
\Sigma^s_\psi=\eta^s_\psi
$ 
& 
$
\displaystyle
\frac{1}{3}
\frac{8N_\Sigma}{\pi^2 \tilde{N} }
$ 
& 
$
\displaystyle
\frac{1}{2}
\frac{\epsilon\ N_\Sigma}
{\tilde{N}+4-N_\Sigma}
$ 
\\
\\
$\displaystyle\Delta^s_\nu$ 
& 
$
\displaystyle
-\frac{8N_\Sigma}
{\pi^2 \tilde{N} }
$ 
& 
$
\displaystyle
-\frac{\epsilon \  N_\Sigma}
{\tilde{N} +4-N_\Sigma}
$ 
\\
\\
$\displaystyle \Delta^s_\rho$ & 
$\displaystyle 
\frac{8N_\Sigma}{\pi^2 \tilde{N} }
$ 
& 
$\displaystyle 
\frac{\epsilon\ N_\Sigma}{\tilde{N} +4-N_\Sigma}
$ 
\\
\\
$
\displaystyle 
\Delta^s_\nu-\Sigma^s_\psi=\eta^s_\nu
$ 
& 
$
\displaystyle 
\frac{-4}{3} \frac{8N_\Sigma}{ \pi^2\tilde{N}}
$ 
& 
$\displaystyle
\frac{-3}{2} 
\frac{\epsilon\ N_\Sigma}
{\tilde{N}+4-N_\Sigma} 
$ 
\\
\\
$\displaystyle 
\Delta^s_\rho-\Sigma^s_\psi=\eta^s_\rho
$ 
& 
$
\displaystyle 
\frac{2}{3} \frac{8N_\Sigma}{\pi^2\tilde{N}}
$ 
& 
$
\displaystyle 
\frac{1}{2} 
\frac{\epsilon\ N_\Sigma}{\tilde{N} +4-N_\Sigma} 
$ 
\\
\\
$\displaystyle \eta^s_{\bar{\psi}\psi}$ 
& 
$
\displaystyle 
\frac{8}{3}\frac{8 N_\Sigma}{\pi^2 \tilde{N}}
$ 
& 
$
\displaystyle 
 \frac{3 \epsilon\ N_\Sigma}{\tilde{N} +4-N_\Sigma}
$ 
\\
\\
$
\displaystyle \eta^s_{\bar{\psi}\Sigma \psi}$ 
& 
$\displaystyle 
\frac{-4}{3}\frac{8 N_\Sigma}{\pi^2\tilde{N}}
$ 
& 
$
\displaystyle 
-\frac{\epsilon\ N_\Sigma}{\tilde{N}+4-N_\Sigma} 
$ 
\\
\end{tabular}
\end{ruledtabular}
\caption{Anomalous dimensions derived from
the large-$N_f$ and $\epsilon$ expansions.} 
\label{tbll}
\end{table}

With diagramatics in 
Secs.\ \ref{Large-Nf expansion}
and \ref{epsilon-expansion}
in hand,
we can proceed for calculations of anomalous dimensions
using minimal subtraction scheme 
\cite{justinbook}. 
The renormalization conditions are given by
\begin{align}
i\Slash{p} 
\left(Z^s_\psi+\Sigma^s_\psi\ \textbf{Div}
\right) & = finite, 
\\
Z^s_{\psi}\nu_0+\Delta^s_\nu\ \textbf{Div}\ \nu & =\nu, 
\\
Z^s_{\psi}\rho_0+\Delta^s_\rho\ \textbf{Div}\ \rho & =\rho,  
\label{ren cond}
\end{align}
where $s=N_f$ or $s=\epsilon$ for 
the large-$N_f$ and $\epsilon$ expansions;
$\textbf{Div}$ represents the divergence,
$\textbf{Div}=\log\left(\Lambda/k\right)$ 
for the large-$N_f$ expansion 
and 
$\textbf{Div}= 1/\epsilon$ for the $\epsilon$ expansions;
$\nu_0$ and $\rho_0$ are the bare masses. 
We can readily read off
the field renormalization 
$Z^s_{\psi}$ 
for the fermion field
and
its anomalous dimension $\eta_{\psi}$,
as
\begin{align}
Z^s_\psi &= 1-\Sigma^s_\psi \textbf{Div}, 
\nonumber \\
\eta^s_\psi &= \Sigma^s_\psi. 
\end{align}

The dimension $d_{\mathcal{O}}$ of an operator $\mathcal{O}$ 
at a critical point is defined 
from the power-law decay of its correlation function as 
$\langle \mathcal{O}(\textbf{r}) \mathcal{O}(\textbf{0}) \rangle 
\propto |\textbf{r}|^{-2d_\mathcal{O}}$. 
In $d=(2+1)$ dimensions, 
the engineering dimensions of 
the mass terms $\bar{\psi}\psi$ 
and $\psi \Sigma \psi$ are 2, 
$
d^{(0)}_{\psi \psi}=d^{(0)}_{\psi \Sigma \psi}=2
$.
Accordingly, the engineering dimensions of 
the masses $\nu$ and $\rho$ are $1$,
$d^{(0)}_{\bar{\psi}\psi}=d-d^{(0)}_\nu=1$ 
and 
$d^{(0)}_{\bar{\psi}\Sigma\psi}=d-d^{(0)}_\rho=1$. 
\cite{justinbook} 

From the renormalization conditions (\ref{ren cond}),
to leading order in $N_f$ or in $\epsilon$, 
and in $d=(2+1)$ dimensions, 
the anomalous part of the
dimensions of the masses $\nu$ and $\rho$ 
can be read off as 
\begin{align}
\eta^s_\nu &= \Delta_\nu^s-\Sigma_\psi^s, \\
\eta^s_\rho &= \Delta_\rho^s-\Sigma_\psi^s.
\label{anomalous dim}
\end{align}
The total dimensions of the masses $\nu$ and $\rho$ are then given by 
$d^s_\nu=1+\eta^s_\nu$ and $d^s_\rho=1+\eta^s_\rho$.
In turn, 
the dimensions of the fermion bilinears are then given 
by
$d^s_{\bar{\psi}\psi}=d-d^s_\nu$ 
and 
$d^s_{\bar{\psi}\Sigma\psi}=d-d^s_\rho$
with 
the corresponding anomalous dimensions 
given by 
$\eta^s_{\bar{\psi}\psi} = 2d^s_{\bar{\psi}\psi}-2d$ 
and 
$\eta^s_{\bar{\psi}\Sigma\psi} = 2d^s_{\bar{\psi}\Sigma\psi}-2d$.  
These results are summarized in Table \ref{tbll}.

\section{Discussion}
\label{Discussion}

\textit{summary of results} 
Two different order parameters, 
$\vec{v}_1$ and $\vec{v}_2$ 
are said to be dual to each other
when a static defect in one of them 
traps a quantum number (or ``charge'') of the other. 
In this paper,
we have discussed 
the interaction effects on
such a pair of order parameters.
The complementary nature of the pair of the order parameters
shows up also in their dynamical properties (correlation functions)
in the following sense:
When a quantum phase transition is driven by fluctuations 
in one of the order parameter, $\vec{v}_1$, say, 
approaching the transition from the disordered (paramagnetic) side,
the order parameter correlation 
at the critical point is reduced.
On the other hand, such fluctuations enhance 
the correlation of the order parameter $\vec{v}_2$, 
which is dual to $\vec{v}_1$.

If we can experimentally monitor, 
at a given quantum critical point,
the correlation functions of a pair of such order parameters, 
the complementary relationship (duality) can be revealed
by extracting their anomalous dimensions, 
which can, at the same time, be helpful to identify the main driving forces 
of the transition, as discussed above. 
For example, 
for the N\'eel and VBS order parameters
which can compete with each other around a quantum critical point,
the neutron scattering experiment
can detect the N\'eel order, 
whereas the scanning tunneling spectroscopy
measurement
can extract some information on the VBS order.
Another probe which can potentially be used to detect 
the correlation of the VBS order parameter
is its coupling to phonons.
As a consequence of such coupling, 
information on the VBS correlation can 
be extracted by looking at the imaginary part of the phonon response 
function 
\cite{cross} 
using, say, X-ray scattering. 
Similar method has been used to identify 
fluctuations in the dimer order parameter
in spin-chains which go through spin-Peierls transition.
\cite{abel}

We, again, emphasize that 
the type of competitions among order parameters discussed in this paper 
has a topological origin, 
rather than some accidental energetic reasons,
and hence is largely independent of microscopic details of the system.
Whether or not such duality relationship among correlation functions holds 
beyond Dirac systems that we discussed 
in terms of the large-$N_f$ and $\epsilon$ expansions, 
should further be studied in future. 
For example, numerical study in the $J-Q$ model\cite{sandvik} would be 
promising in this regard. 

\textit{description in terms of the NL$\sigma$M} 
To further illustrate the topological origin behind the duality, 
let us seek for an alternative description of fluctuating 
order parameters. 
Instead of taking the non-interacting Dirac fermions as a starting point, consider the system which is deep inside the ordered phase
(the large-$g N_f$ region in Fig.\ \ref{fig: phase dirgram}). 
It is then reasonable to employ a description in terms of
the fluctuating but non-vanishing order parameter. 
The dynamics of the order parameter
may be described by  
the $\mathrm{O}(N_{\Sigma})$ NL$\sigma$M
whose kinetic term is given by,
the following imaginary time action 
\begin{eqnarray}\label{skin}
  S_{\mathrm{NL} \sigma \mathrm{M}}
=
\frac{1}{t}
\int d^3x\,
\partial_{\mu} n_a 
\partial_{\mu} n_a 
\end{eqnarray}
where $\sum_{a=1}^{N_{\Sigma}} n^2_{a} =1$
[see Eq.\ (\ref{nvector})], and 
$t$ is the coupling constant of the NL$\sigma$M. 
Such NL$\sigma$M description can be derived,
inside the ordered phase, by integrating over gapped 
fermions in the presence of slowly varying order parameter 
field $\vec{v}_1$\cite{Tanaka-Hu05}.
The NL$\sigma$M is in the weak coupling regime (small $t$)
when $g$ 
(in Eq.\ \ref{eq: generic Hamiltonian}, say) is large, while it is in the strong coupling regime
(large $t$)
when $g$ approaches the critical coupling $g_*$ from the disordered side.

In addition to the NL$\sigma$M kinetic term, the integration
over fermions gives rise to either
the Wess-Zumino-Witten (WZW)-term,
$\theta$-term,
or the Berry phase term
in the NL$\sigma$M 
(for alternative derivation of topological terms see Ref. \onlinecite{hong}).
For example, when we are in the $\mathrm{O}(4)$ ordered phase,
the NL$\sigma$M is supplemented with the $\theta$-term, 
\begin{eqnarray}
S_{\theta}
=
{i} 
\frac{\theta}{2\pi^2}  
\int d^3x\,
\epsilon_{abcd} \  n_a \partial_\tau n_b \partial_x n_c \partial_y n_d.  
\end{eqnarray}
Here, 
the $\theta$ angle can be computed
from the microscopic Dirac model
\cite{Tanaka-Hu05},
\begin{align}
\frac{\theta}{\pi} =
1 - \frac{9}{8} \cos \varphi
+ \frac{1}{8} \cos 3\varphi, 
\label{theta}
\end{align}
where $\varphi$ is determined
from the ratio between the amplitude of the 
$\mathrm{O}(4)$ order parameter (as defined in Sec.\ \ref{ffermion})
and the mass term,
$\tan \varphi =-|\boldsymbol{v}|/m_3$:
When fermions are gapped solely because of 
the $\mathrm{O}(4)$ order parameter (i.e. $m_3=0$)
$\boldsymbol{v}=|\boldsymbol{v}|\boldsymbol{n}$, 
which develops a finite
expectation value in the $\mathrm{O}(4)$ ordered phase, 
$\theta=\pi$.
On the other hand, for example, 
if we add a mass term 
$m_{3}\bar{\psi} \Sigma_3 \psi$ in the ordered phase, 
which anticommutes with the $\mathrm{O}(4)$ order parameter 
$v_a\bar{\psi} \Sigma_a \psi$,
the $\theta$-term deviates from $\theta=\pi$.

Following the spirit of the Goldstone-Wilczek formula
\cite{Goldstone81},
when the order parameters are treated as a static background, 
we can see that it is the $\theta$-term which implements the duality 
within the NL$\sigma$M field theory. 
When this operator acquires an expectation value, the VBS order results.
In this manner the NL$\sigma$M field theory contains 
the ingredients describing 
both the ordered and paramagnetic phases.
This is in fact in line with the known fact that 
the Berry phase term in the (2+1) dimensional 
O(3) NL$\sigma$M secretly encodes the VBS order, 
and that
in the paramagnetic phase the presence of the Berry phases leads to
the VBS order.
\cite{SenthilScience,Senthiletal2004}

\textit{speculation on the criticality and 
the RG flow of the NL$\sigma$M with $\theta$ term}

Having established the duality relation within 
the ordered phase of the NL$\sigma$M,
it is now interesting to ask what would happen 
when we approach the strongly coupled region 
where the order parameter fluctuations are strong.
In particular, 
how does the RG flow of the $\mathrm{O}(4)$ NL$\sigma$M
with $\theta$-term look like? 
The relevance of the topological terms
($\theta$-term as well as other topological terms such as the
Wess-Zumino-Witten term)
in the NL$\sigma$M 
to the deconfined quantum criticality has been discussed in 
Refs. \onlinecite{Tanaka-Hu05, Tanaka-Hu06, Senthil-Fisher05}.
(See also a recent a paper \onlinecite{Xu2011}).

Unfortunately, the role of the topological term
on the quantum criticality in (2+1)-dimensions 
is not well-understood.
This situation should be contrasted with 
the state of our understanding on 
the $\theta$-term in the O(3) NL$\sigma$M in (1+1)-dimensions; 
Starting from the Haldane conjecture, 
it is by now well-known that the different nature of the ground
states of the $S=\mbox{integer}$ and $S=\mbox{half-odd integer}$
spin chain is reflected to the value of the $\theta$-term,
$\theta = 2\pi \times S$ (mod $2\pi$) in the O(3) NL$\sigma$M. 
Quite surprisingly,
in (1+1)-dimensions, 
the integrability of the O(3) NL$\sigma$M
allows us to ``prove'' that 
the O(3) NL$\sigma$M at $\theta=\pi$ 
flows, in the RG sense,  
into a gapless critical point
(the SU(2) Wess-Zumino-Witten theory at level one). 
The low-lying excitation at the critical point consists,
not of magnons,
but (fermionic) spinons. 
When detuned from $\theta=\pi$ 
(by breaking a link parity symmetry, say,
with bond dimerization in the Heisenberg exchange coupling), 
the $\theta $-term flows to $0$ or $2\pi$, where a system is in a gapped 
paramagnetic phase.

We now present some speculations 
on the effect of the topological term 
in (2+1) dimensions 
in the region when the fluctuations of the 
NL$\sigma$M order parameter is strong.
Let us
start from the ordered phase of the NL$\sigma$M;
it exists for the small NL$\sigma$M coupling, $t\ll 1$,
and this should be true for any value of $\theta$. 
As we crank up the NL$\sigma$M coupling constant $t$, 
the interactions between ``magnons'' become stronger.
It would then seem reasonable to assume that,
even in the presence of non-vanishing $\theta$,  
we reach a critical point 
by eventually destroying the (long-range) order,
at which the NL$\sigma$M order parameter develops
a critical (power-law) correlation
with vanishing expectation value, $\vec{v}=0$. 

Recall that
in the conventional NL$\sigma$M without any topological term,
the nature of such transition 
can be captured by e.g. the $2+\epsilon$ expansion.
On the other hand if we stay exactly in three space-time dimensions, 
which is necessary if we are after the effect of the topological term,
the NL$\sigma$M is not perturbatively renormalizable
(although it may be renormalizable in terms of the non-perturbative RG).
It should the be considered as an effective field theory,
and is not UV complete. 
As we approach the putative quantum critical region 
(the strong coupling region of the NL$\sigma$M) 
described above,
the NL$\sigma$M description should be replaced by
some other description. 
The question then is, to which extent 
we can deduce such UV (high-energy) description 
for the case of non-vanishing $\theta$.

In terms of the corresponding fermionic model Sec.\ \ref{ffermion},
which should be regarded as a more microscopic (or UV) description, 
increasing $t$ (the coupling constant of the NL$\sigma$M)
corresponds to reducing the four fermion coupling constant $g$.
In terms of the fermionic phase diagram 
(Fig.\ \ref{fig: phase dirgram}),
we thus approach the phase boundary 
from the ordered side.
In particular, $|\boldsymbol{v}|=0$ 
 corresponds to the case of $\theta=0$,
and in this case, 
we approach the GN critical point separating 
the ordered phase and paramagnetic phase (Dirac semi-metal phase).
One would then conclude that 
a UV critical point of the NL$\sigma$M at $\theta=\pi$,
if exists, 
would correspond to the GN fixed point.

This is a tempting, but of course very dangerous, argument.
We would, however, expect,
by matching with the fermionic description, 
the following: 
at the UV critical point of the NL$\sigma$M at $\theta=\pi$, 
not only the NL$\sigma$M order parameter 
develops critical correlation,
but its dual order parameter also does so. 
(These order parameters are called $\vec{v}_1$ and $\vec{v}_2$ 
in the beginning of this section). 
This is, of course, a complete surprise from 
the NL$\sigma$M point of view:
There is no inkling of $\vec{v}_2$ order parameter
what so ever in the ordered phase,
and controlling coupling constant of the NL$\sigma$M 
($t$ in Eq.\ \ref{skin}) has nothing to do with $\vec{v}_2$ order. 
Nevertheless because of the $\theta$-term, 
making $t$ large (destabilizing the order parameter $\vec{v}_1$)
somehow makes $\vec{v}_2$ correlation stronger,
since 
the $\vec{v}_2$ order parameter within the NL$\sigma$M
is realized as a defect in the $\vec{v}_1$ parameter.
This is in fact in line with the non-trivial identification 
of the VBS order parameter as the skyrmion creation operator
in
the (2+1) dimensional O(3) NL$\sigma$M augmented with
the Berry phase term.
\cite{SenthilScience,Senthiletal2004}
The VBS phase arises as a proliferation of monopoles 
in the presence of non-trivial Berry phases. 
Thus, while the more precise nature of the putative quantum critical
point at $\theta=\pi$ is difficult to study, 
overall physical picture related to the physics of 
$\theta$-term seems to be inferred
from the fermionic GN model description.

Another interesting issue is 
the RG flow which incorporates both $t$ and $\theta$ 
(the couplings of kinetic and topological terms in the NL$\sigma$M). 
In the weakly coupled region ($t/\Lambda \ll 1$), 
the presence of the non-vanishing $\theta$ 
little affects the RG flow,
while this may not be the case in the strongly coupled
region ($t/\Lambda  \simeq 1$).
In fact, this is the case for the NL$\sigma$M on
the Grassmannian manifold in two dimensions, which
is relevant to the two parameter scaling flow of the 
quantum Hall effect. 
\cite{QHEbook}
Again by matching the NL$\sigma$M description
with the fermionic GN model description,
since the $\theta$-term is correlated to the mass term
$m_b\bar{\psi}\Sigma_b\psi$
(see Eq.\ \ref{theta}), 
it would be then 
interesting to ``guess'' the RG 
flow of the $\theta$-term
from the RG flow of the mass term. 
Within the large-$N_f$ or $\epsilon$ expansion,
the scaling dimension of the fermion mass $m_b\bar{\psi}\Sigma_b\psi$
at the GN critical point is very close to the scaling dimension
at the trivial (non-interacting Dirac) fixed point.
Thus, when $1/N_f$ or $\epsilon$ is small enough,
the mass term is a relevant perturbation to the GN fixed point,
and it grows under the RG. 
This would suggest 
the $\theta$-term is also relevant in the strong coupling
region of the NL$\sigma$M at $\theta=\pi$;
the deviation of the $\theta$-angle from $\theta=\pi$
grows under the RG. 
However, in principle, more exotic possibility,
i.e., the $\theta$-term at the non-trivial critical point being irrelevant,
can also be realized. 
We leave these issues for future studies.

\section*{acknowledgments}

We would like to thank D. H. Lee for insightful discussions. 
SR thanks the Center for Condensed Matter Theory at University of California,
Berkeley for its support. PG acknowledges funding from LBNL DOE-504108.

\appendix

\bibliography{draft_large_n_2012_01_03}

\end{document}